\input psfig.sty
\documentstyle[12pt]{article}
\def\beq{\begin{equation}}
\def\eeq{\end{equation}}
\def\beqn{\begin{eqnarray}}
\def\eeqn{\end{eqnarray}}
\def\twiddle{\lower.9ex\rlap{$\kern-.1em\scriptstyle\sim$}}
\def\bigtwiddle{\lower1.ex\rlap{$\sim$}}
\def\gtwid{\mathrel{\raise.3ex\hbox{$>$\kern-.75em\lower1ex\hbox{
$\sim$}}}}
\def\ltwid{\mathrel{\raise.3ex\hbox{$<$\kern-.75em\lower1ex\hbox{
$\sim$}}}}

\begin{document}
\title{Dark Matter Axions and Caustic Rings}
\author{Pierre Sikivie\\
{\it Department of Physics}\\
{\it University of Florida}\\
{\it Gainesville, FL  32611}}
\date{}
\maketitle

\vspace{.2in}
\section*{Contents:}

\begin{enumerate}
\item The strong CP problem
\item Dark matter axions
\item The cavity detector of galactic halo axions
\item Caustic rings in the density distribution of cold dark matter halos
\end{enumerate}

\section{The strong CP problem}

The axion was postulated two decades ago \cite{arev} to explain 
why the strong interactions conserve $P$ and $CP$ in spite of the fact 
that the standard model as a whole violates those symmetries.  Consider
the Lagrangian of QCD:
\beqn
{\cal L}_{QCD} &=& -{1\over 4} G^a_{\mu\nu} G^{a\mu\nu} +\sum_{j=1}^n
\left[ \overline q_j \gamma^\mu i D_\mu q_j - (m_j q^+_{Lj} q_{Rj}
+ \hbox{h.c.})\right]\nonumber\\
& & + {\theta g^2\over 32\pi^2} G^a_{\mu\nu} \tilde G^{a\mu\nu} \,\,\,  .
\eeqn
The last term is a 4-divergence and hence does not contribute in 
perturbation theory.  That term does however contribute through 
non-perturbative effects \cite{thft} associated with QCD instantons 
\cite{inst}.  Such effects can make the physics of QCD depend upon the 
value of $\theta$. Using the Adler-Bell-Jackiw anomaly \cite{abj}, one 
can show that $\theta$ dependence must be present if none of the current 
quark masses vanishes.  (If this $\theta$ dependence were absent,
QCD would have a $U_A(1)$ symmetry and would predict the mass of 
the $\eta'$ pseudo-scalar meson to be less than 
$\sqrt{3} m_\pi \approx 240$~MeV \cite{SW},contrary to observation.)  
One can further show that QCD depends upon $\theta$ only through the 
combination of parameters:
\beq
\overline \theta =\theta -\hbox{arg} (m_1, m_2, \ldots m_n)
\eeq
If $\overline \theta \neq 0$, QCD violates $P$ and $CP$.  The absence of
$P$ and $CP$ violations in the strong interactions therefore places an
upper limit upon $\overline\theta$.  The best constraint follows from 
the experimental bound \cite{ned} on the neutron electric dipole moment 
which yields: $\overline\theta < 10^{-9}$.

The question then arises:  why is $\overline\theta$ so small?  In the 
standard model of particle physics, the quark masses originate in 
the electroweak sector of the theory.  This sector must violate $P$ and 
$CP$ to produce the correct weak interaction phenomenology.  There is 
no reason in the standard model to expect the overall phase of the quark 
mass matrix to exactly match the value of $\theta$ from the QCD sector so 
that $\overline \theta < 10^{-9}$. In particular, if $CP$ violation is 
introduced in the manner of Kobayashi and Maskawa \cite{KM}, the Yukawa 
couplings that give masses to the quarks are arbitrary complex numbers 
and hence arg~det~$m_q$ and $\overline\theta$ have no reason to take on 
any special value at all.

The problem why $\overline\theta < 10^{-9}$ is usually referred to as the 
``strong $CP$ problem''.  The existence of an axion would solve it.  There 
are other solutions.  Setting $m_u = 0$ removes the $\theta$-dependence 
of QCD and hence the strong $CP$ problem as well.  
However, $m_u = 0$ may cause problems with the successful current algebra 
relations among pseudo-scalar meson masses.  I refer the reader to 
refs.\cite{NS,Leut} for recent discussions of the issues involved.  Another 
type of solution involves the assumption that $CP$ and/or $P$ is 
spontaneously broken but is otherwise a good symmetry.  In this case, 
$\overline\theta$ is calculable and may be arranged to be small \cite {CPsb}. 
Finally, let's emphasize that the strong $CP$ problem need not be solved 
in the low energy theory.  Indeed, as Ellis and Gaillard \cite{EG} 
pointed out, if in the standard model $\overline\theta=0$ near the 
Planck scale, then $\overline\theta \ll 10^{-9}$ at the QCD scale.

Peccei and Quinn \cite{PQ} proposed to solve the strong $CP$ problem by 
postulating the existence of a global $U_{PQ}(1)$ quasi-symmetry.  To do 
its job, $U_{PQ}(1)$ must be a symmetry of the theory at the classical 
(i.e., at the Lagrangian) level, it must be broken explicitly by those 
non-perturbative effects that make the physics of QCD depend upon $\theta$, 
and finally it must be spontaneously broken.  The axion \cite{WW} is the 
quasi-Nambu-Goldstone boson associated with the spontaneous breakdown 
of $U_{PQ}(1)$. One can show that, if a $U_{PQ}(1)$ quasi-symmetry is 
present, then
\beq
\overline\theta = \theta - arg (m_1 \ldots m_n) - {a(x)\over f_a}\, ,
\eeq
where $a(x)$ is the axion field and $f_a$, called the axion decay 
constant, is of order the vacuum expectation value (VEV) which spontaneously
breaks $U_{PQ}(1)$. \ It can further be shown \cite{VW} that the 
non-perturbative effects that make QCD depend upon $\overline\theta$ 
produce an effective potential $V(\overline\theta)$ whose minimum is 
at $\overline\theta =0$.\ \ Thus, by postulating an axion, $\overline\theta$ 
is allowed to relax to zero dynamically and the strong $CP$ problem is solved.

The properties of the axion can be derived using the methods of current
algebra \cite{WW,curr}.  The axion mass is given in terms of $f_a$ by
\beq 
m_a\simeq 0.6~eV~{10^7 GeV\over f_a}\, .
\eeq
All the axion couplings are inversely proportional to $f_a$.  For example,
the axion coupling to two photons is:
\beq
{\cal L}_{arr} = -g_\gamma {\alpha\over \pi} {a(x)\over f_a}
\vec E \cdot\vec B
\eeq
where $\vec E$ and $\vec B$ are the electric and magnetic fields, 
$\alpha$ is the fine structure constant, and $g_\gamma$ is a model-dependent 
coefficient of order one.  $g_\gamma=0.36$ in the DFSZ model \cite{DFSZ}
whereas $g_\gamma=-0.97$ in the KSVZ model \cite{KSVZ}.  A priori the 
value of $f_a$, and hence that of $m_a$, is arbitrary.  However, searches 
for the axion in high energy and nuclear physics experiments combined with
astrophysical constraints, the latter derived by considering the effect
of the axion upon the lifetimes of red giants and SN1987a, rule out 
$m_a \gtwid 3~10^{-3}$~eV \cite{arev,jk}. \ In addition, as will be
discussed 
in section~II, cosmology places a lower limit on $m_a$ of order
$10^{-6}$~eV by requiring that axions do not overclose the universe.

\section{Dark matter axions}
For small masses, axion production in the early universe is dominated
by a novel mechanism \cite{ac}.  The crucial point is that the 
non-perturbative QCD effects that produce the effective potential 
$V(\overline\theta)$ are strongly suppressed at temperatures high 
compared to $\Lambda_{QCD}$ \cite{GPY}. At these high temperatures, 
the axion is massless and all values of $\langle a(x)\rangle$ are 
equally likely.  At $T\simeq 1$~GeV, the potential $V$ turns on and 
the axion field starts to oscillate about a $CP$ conserving minimum 
of $V$. These oscillations do not dissipate into other forms of energy 
because, in the relevant mass range, the axion is too weakly coupled for 
that to happen.  The oscillations of the axion field may be described as 
a fluid of axions.  The typical momentum of the axions in the fluid is 
the inverse of the correlation length of the axion field.  Because that 
correlation length is of order the horizon, we have 
$p_a \sim (10^{-6}$ sec)$^{-1} \sim 10^{-9}$~eV at $T\simeq 1$~GeV.
Afterwards, $p_a$ decreases in time as $R^{-1}$ where $R$ is the
cosmological scale factor.  Thus the axion fluid is very cold compared 
to the ambient temperature.

Let me briefly indicate how the present cosmological energy density of this
axion fluid is estimated.  Let $\varphi(x)$ be the complex scalar field 
whose VEV $v$ spontaneously breaks $U_{PQ}(1)$. \ At extremely high 
temperatures, the $U_{PQ}(1)$ symmetry is restored.  It becomes spontaneously 
broken when the temperature drops below a critical value $T_{PQ}$ of order 
$v$. Below $T_{PQ}$, the axion field $a(x)$ appears as the phase of the 
$VEV$ of $\varphi$:  $\langle \varphi(x) \rangle = ve^{ia(x)/v}$.  
We must now distinguish two cases.  Either inflation occurs with reheat 
temperature below $T_{PQ}$, or not (i.e., inflation does not occur or it 
occurs with reheat temperature above $T_{PQ}$).  In the first case, inflation 
homogenizes the axion field and there is only one contribution to the axion 
cosmological energy density, the contribution from so-called ``vacuum 
misalignment''.  In the second case, there are additional contributions 
from axion string and axion domain wall decay.  Only the contribution from 
vacuum misalignment is  discussed in any detail here.

When the axion mass turns on near the QCD phase transition, the axion
field starts to oscillate about one of the $CP$ conserving minima of
the effective potential.  The oscillation begins approximately at
cosmological time $t_1$ such that $t_1m_a (T(t_1)) = 0(1)$, where 
$m_a(T)$ is the temperature dependent axion mass.  Soon after $t_1$,
the axion mass changes sufficiently slowly that the total number of
axions in the oscillations of the axion field is an adiabatic
invariant.  $T_1 \equiv T(t_1)$ has been estimated to be of order 
1~GeV.\ \ The number density of axions at time $t_1$ is
\beq
n_a(t_1)\simeq {1\over 2} m_a(t_1) \langle a^2(t_1)\rangle \simeq
\pi f_a^2 {1\over t_1}
\eeq
where $f_a = {v\over N}$ is the axion decay constant introduced 
earlier.  $N$ is an integer which expresses the color anomaly of
$U_{PQ}(1)$. \ $N$ also equals the number of $CP$ conserving vacua \cite{adw}
at the bottom of the 'Mexican hat' potential, i.e., in the interval 
$0\leq {a\over v} < 2\pi$.  In Eq.~(6), we used the fact that the 
axion field $a(x)$ is approximately homogeneous on the horizon scale 
$t_1$.  Wiggles in $a(x)$ which entered the horizon long before $t_1$ 
have been red-shifted away \cite{Vil}.  We also used the fact 
that the initial departure of $a(x)$ from the nearest minimum is of 
order ${v\over N} = f_a$. \ The axions of Eq.~(6) are decoupled and
non-relativistic.  Assuming that the ratio of the axion number 
density to the entropy density is constant from time $t_1$ till
today, one finds \cite{ac}
\beq
\Omega_a = \left({0.6~10^{-5}\hbox{\ eV}\over m_a}\right)^{7\over 6}
\left({200\hbox{\ MeV}\over \Lambda_{QCD}}\right)^{3\over 4}
\left({75\hbox{\ km/s}\cdot \hbox{Mpc}\over H_0}\right)^2
\eeq
for the ratio of the axion energy density to the critical density
for closing the universe.  $H_0$ is the present Hubble rate.
Eq.~(7) implies the bound $m_a \gtwid 10^{-6}$~eV. 

However, there are many sources of uncertainty in the estimate of
$\Omega_a$. \ The axion energy density may be diluted by the entropy
release from heavy particles which decouple before the QCD epoch but decay
afterwards \cite{ST}, or by the entropy release associated with a first
order QCD phase transition.  On the other hand, if the QCD phase
transition is first order \cite{pt}, an abrupt change of the axion mass at
the transition may increase $\Omega_a$.  If inflation occurs with reheat
temperature less than $T_{PQ}$, there may be an accidental suppression of
$\Omega_a$ because the homogenized axion field happens to lie close to a
$CP$ conserving minimum.  Because the RHS of Eq.~(7) is multiplied in
this case by a factor of order the square of the initial vacuum
misalignment angle ${a(t_1)\over v}N$ which is randomly chosen between
$-\pi$ and $+\pi$, the probability that $\Omega_a$ is suppressed by a
factor $x$ is of order $\sqrt{x}$.  This rule cannot be extended to
arbitrarily small $x$ however because quantum mechanical fluctuations in
the axion field during the epoch of inflation do not allow the suppression
to be perfect \cite{inflax}.  If inflation occurs with reheating
temperature larger than $T_{PQ}$ or if there is no inflation, there are
contributions to $\Omega_a$ from axion string \cite{rd} and axion domain
wall decay \cite{add} in addition to the contribution, Eq.~(7), from
vacuum misalignment.  My collaborators and I \cite{ssa} have estimated
each of these additional contributions to be of the same order of
magnitude as that from vacuum misalignment.  Other authors \cite{rd,osa}
have estimated that the contribution from axion string decay dominates
over that from vacuum misalignment by a factor 100 or a factor 10. 

The axions produced when the axion mass turns on during the QCD phase
transition are cold dark matter (CDM) because the axions are
non-relativistic from the moment of their first appearance at 1~GeV
temperature.  Studies of large scale structure formation support the view
that the dominant fraction of dark matter is CDM \cite{MST}.  Moreover
any form of CDM necessarily contributes to galactic halos by falling into
the gravitational wells of galaxies.  Hence, there is excellent motivation
to look for CDM candidates as constituent particles of our galactic halo,
even after some fraction of our halo has been demonstrated to be in MACHOs
\cite{MACHO} or some other form. 

Finally, let's mention a particular kind of clumpiness \cite{amc} which 
affects axion dark matter if there is no inflation after the Peccei-Quinn 
phase transition.  This is due to the fact that the dark
matter axions are inhomogeneous with $\delta \rho / \rho \sim 1$ over the
horizon scale at temperature $T_1 \simeq$ 1 GeV, when they are produced at
the start of the QCD phase-transition, combined with the fact that their
velocities are so small that they do not erase these inhomogeneities by
free-streaming before the time $t_{eq}$ of equality between the matter and
radiation energy densities when matter perturbations can start to grow. 
These particular inhomogeneities in the axion dark matter are immediately
in the non-linear regime after time $t_{eq}$ and thus form clumps, called
`axion mini-clusters' \cite{amc}.  These have mass $M_{mc} \simeq 10^{-13}
M_\odot$ and size $l_{mc} \simeq 10^{12}$ cm.  

\section{The cavity detector of galactic halo axions} Axions can be
detected by stimulating their conversion to photons in a strong magnetic
field \cite{ps}.  The relevant coupling is given in Eq.~(5).  In
particular, an electromagnetic cavity permeated by a strong static
magnetic field can be used to detect galactic halo axions.  The latter
have velocities $\beta$ of order $10^{-3}$ and hence their energies
$E_a=m_a+{1\over 2} m_a\beta^2$ have a spread of order $10^{-6}$ above the
axion mass.  When the frequency $\omega=2\pi f$ of a cavity mode equals
$m_a$, galactic halo axions convert resonantly into quanta of excitation
(photons) of that cavity mode.  The power from axion $\to$ photon
conversion on resonance is found to be \cite{ps,kal}:  \beqn P &=& \left
({\alpha\over\pi} {g_\gamma\over f_a}\right )^2 V\, B_0^2 \rho_a C {1\over
m_a} \hbox{Min}(Q_L,Q_a)\nonumber\\ &=& 0.5\; 10^{-26} \hbox{Watt}\left(
{V\over 500\hbox{\ liter}} \right) \left({B_0\over 7\hbox{\
Tesla}}\right)^2 C\left({g_\gamma\over 0.36}\right)^2 \nonumber\\ &\cdot&
\left({\rho_a\over {1\over 2} \cdot 10^{-24} {{\scriptstyle g_r}\over
\hbox{\scriptsize cm}^3}}\right) \left({m_a\over 2\pi (\hbox{GHz})}\right)
\hbox{Min}(Q_L,Q_a)  \eeqn where $V$ is the volume of the cavity, $B_0$ is
the magnetic field strength, $Q_L$ is its loaded quality factor,
$Q_a=10^6$ is the `quality factor' of the galactic halo axion signal (i.e.
the ratio of their energy to their energy spread), $\rho_a$ is the density
of galactic halo axions on Earth, and $C$ is a mode dependent form factor
given by \beq C = {\left| \int_V d^3 x \vec E_\omega \cdot \vec
B_0\right|^2 \over B_0^2 V \int_V d^3x \epsilon |\vec E_\omega|^2} \, \eeq
where $\vec B_0(\vec x)$ is the static magnetic field, $\vec E_\omega(\vec
x) e^{i\omega t}$ is the oscillating electric field and $\epsilon$ is the
dielectric constant. 

Because the axion mass is only known in order of magnitude at
best, the cavity must be tunable and a large range of frequencies
must be explored seeking a signal.  The cavity can be tuned by
moving a dielectric rod or metal post inside it.  Using Eq.~(8), one 
finds that to perform a search with signal to noise ratio $s/n$, the 
scanning rate is:
\beqn
{df\over dt} &=& {12 \hbox{GHz}\over \hbox{year}} \left({4n\over s}
\right)^2 \left({V\over 500\hbox{\ liter}}\right)^2
\left( {B_0\over 7\hbox{\ Tesla}}\right)^4 C^2\left({g_\gamma\over
0.36}\right)^4  \nonumber\\
&\cdot &\left({\rho_a\over {1\over 2}\cdot 10^{-24} 
{{\scriptstyle gr}\over \hbox{\scriptsize cm}^3}}\right)^2 
\left({3K\over T_n}\right)^2 \left(
{f\over \hbox{GHz}}\right)^2 {Q_L\over Q_a} \,\,\,  ,
\eeqn
where $T_n$ is the sum of the physical temperature of the cavity plus 
the noise temperature of the microwave receiver that detects the photons 
from $a \to \gamma$ conversion.  Eq.~(10) assumes that $Q_L < Q_a$ and 
that some strategies have been followed which optimize the search rate.
The best quality factors attainable at present, using oxygen
free copper, are of order $10^5$ in the GHz range. To make 
the cavity of superconducting material is probably not useful 
since it is permeated by a strong magnetic field in the experiment.

\begin{figure}
\psfig{file=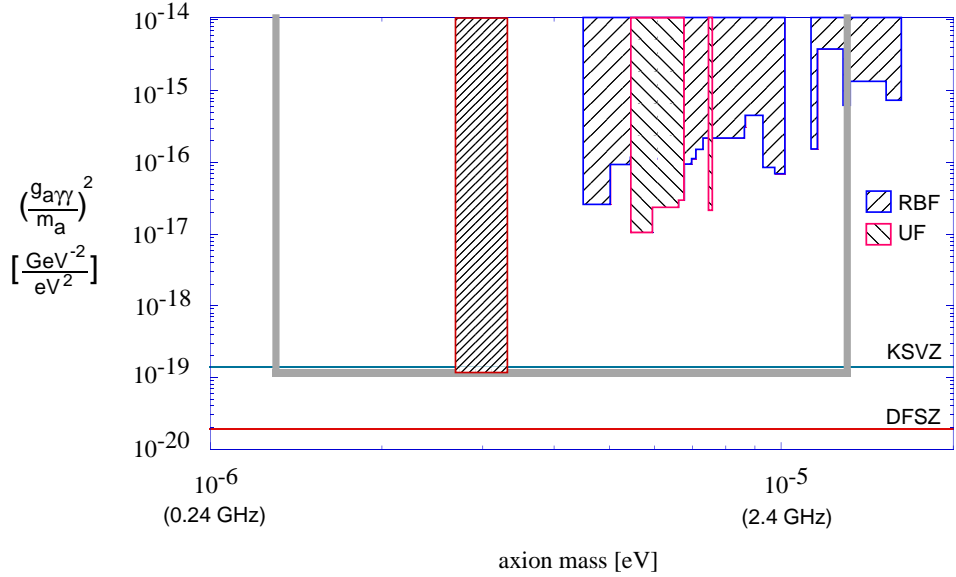,width=14cm}
\caption{Regions in mass - (coupling/mass)$^2$ space which have been 
ruled out by the RBF and UF experiments (hatched), which the LLNL 
experiment has searched so far (densely hatched) and which it expects
to rule out in the future (shaded border).  $g_{a\gamma\gamma} \equiv
{\alpha\over\pi} {g_\gamma\over f_a}$. The DFSZ and KSVZ model 
predictions are also shown.}
\end{figure}

Eq. (8) shows that a galactic halo search at the required
sensitivity is feasible with presently available technology,
provided the form factor $C$ can be kept at values of order one
for a wide range of frequencies.  For a cylindrical cavity and a 
longitudinal magnetic field, $C=0.69$ for the lowest TM mode.
The form factors of the other modes are much smaller.  The 
resonant frequency of the lowest TM mode of a cylindrical cavity
is $f$=115 MHz $\left( {1m\over R}\right)$ where $R$ is the radius
of the cavity.  Since $10^{-6}\hbox{\ eV} = 2\pi$ (242 MHz), a
large cylindrical cavity is convenient for searching the low
frequency end of the range of interest.  To extend the search
to high frequencies without sacrifice in volume, one may
power-combine many identical cavities which fill up the
available volume inside a magnet's bore \cite{rsci,hag}.  This 
method allows one to maintain $C=0(1)$ at high frequencies, albeit 
at the cost of increasing engineering complexity as the frequency, 
and hence the number of cavities, is increased.

Pilot experiments were carried out at Brookhaven National
Laboratory \cite{RBF} and at the University of Florida \cite{UF}
which demonstrated the feasibility of the cavity detection method.  
The (magnetic field)$^2 \times$ volume provided by the magnets used 
in these experiments were relatively low:  $B_0^2V=0.36T^2m^3$ and 
$0.45 T^2 m^3$ respectively for Brookhaven (RBF) and Florida (UF).  
Fig.~1 shows the limits that these experiments placed on the coupling 
$g_{a\gamma\gamma} = {\alpha\over\pi} {g_\gamma\over f_a}$ as a 
function of the axion mass $m_a$ assuming that the local density of 
galactic halo axions is
$\rho_a ={1\over 2}\cdot 10^{-24} gr/\hbox{cm}^3$.  

Second generation experiments are presently under way at Lawrence
Livermore National Laboratory (LLNL)\cite{LLNL} and at Kyoto University
\cite{Kyoto}.  The LLNL experiment uses a much larger magnet ($B_0^2 V =
12 T^2m^3$) than the pilot experiments.  It started taking data in
Feb.~'96.  Up till now, it has searched the range of frequencies 670 
to 800 MHz, or 2.77 to 3.30 $\mu$eV (see Fig. 1), at a level of
sensitivity sufficient to discover KSVZ axions if they are the
constituents of our halo.  The LLNL experiment plans to cover the
mass range $1.3 < m_a < 13 \mu$eV. For the high frequency end of this
range, it will use an array of four cavities kept in tune and whose
outputs are power-combined.  SQUIDs are being developped by members of 
the collaboration as front-end detectors of the microwave power from 
axion to photon conversion.  So far, the LLNL experiment has used the
same microwave detection technology, HEMT amplifiers, as the pilot
experiments. SQUIDs may boost the sensitivity by a factor of ten or more. 
This would allow the LLNL experiment to search for DFSZ axions whereas
only sensitivity to KSVZ axions had been originally planned.

The Kyoto experiment has a magnet of size similar to that of the pilot
experiments but uses a beam of Rydberg atoms to count the photons from
$a\to \gamma$ conversion.  Single photon counting constitutes a dramatic
improvement in microwave detection sensitivity.  With HEMT amplifiers one
needs to have thousands of $a\to \gamma$ conversions per second and
integrate for about 100~sec to find a signal in the noise.  With single
photon counting, a few $a\to \gamma$ conversions suffice in principle.  To
build a beam of Rydberg atoms capable of single photon counting is a
considerable achievement in itself.  In addition, the cavity will be
cooled by a dilution refrigerator down to a temperature ($\sim 10$~mK) 
where the thermal photon background is negligible.  The Kyoto experiment
will first search near $m_a=10^{-5}$~eV. \ \ Its projected sensitivity is
sufficient to discover DFSZ axions even if their local density is only
$1\over 5$ of the local halo density.  

\section{Caustic rings in the density distribution of cold dark 
matter halos}

If a signal is found in the cavity detector of galactic halo axions, it
will be possible to measure their energy spectrum with great precision and
resolution.  Hence there is good motivation to ask what can be learned
about our galaxy from analyzing the signal.  Conversely, if the energy
spectrum of cold dark matter particles on Earth can be deduced from
studies of galactic halo formation, that information may be used to
increase the signal/noise ratio of the search.  It may also be useful
in searching for WIMPs.  Indeed, because they have moved purely under 
the influence of gravity since their decoupling, all cold dark matter
particles have exactly the same phase space distribution and hence the 
same energy spectrum on Earth in the limit where their, at any rate tiny,
primordial velocity dispersions are neglected. 

In many past discussions of dark matter detection on Earth, 
it has been assumed that the dark matter particles have an 
isothermal distribution. Thermalization has been argued to be
the result of a period of "violent relaxation'' following the 
collapse of the protogalaxy. If it is strictly true that the 
velocity distribution of dark matter particles is isothermal, 
which seems to be a strong assumption, then the only 
information that can be gained from its observation is the 
corresponding virial velocity and our own velocity relative to 
its standard of rest. If, on the other hand, thermalization 
is incomplete, a signal in a dark matter detector may yield 
additional information.

J.R. lpser and I discussed \cite{jri} the extent to which the 
phase-space distribution of cold dark matter particles is thermalized 
in a galactic halo and concluded that there are substantial deviations
from a thermal distribution in that the highest energy particles 
have discrete values of velocity.  There is one velocity peak on 
Earth due to dark matter particles falling onto the galaxy for the
first time, one peak due to particles falling out of the galaxy 
for the first time, one peak due to particles falling into the 
galaxy for the second time, etc.  The peaks due to particles that 
have fallen in and out of the galaxy a large number of times in the 
past are washed out because of scattering in the gravitational 
wells of stars, globular clusters and large molecular clouds. But 
the peaks due to particles which have fallen in and out of the 
galaxy only a small number of times in the past are not washed out.

I. Tkachev, Y. Wang and I have used the self-similar infall model of
galactic halo formation to estimate the local densities and the velocity
magnitudes of the dark matter particles in the velocity peaks \cite{stw}. 
We generalized the existing version of that model to take account of the
angular momentum of the dark matter particles.  In the absence of angular
momentum, the model produces flat rotation curves for a large range of
values of a parameter $\epsilon$.  We find that the presence of angular
momentum produces an effective core radius, i.e., it makes the
contribution of the halo to the rotation curve go to zero at zero radius. 
The presence of a core radius is consistent with observation.  The model
provides a detailed description of the large scale properties of galactic
halos including their density profiles, their extent and their total mass. 
By fitting the model to our galactic halo, we obtained the average values
of the local densities and the velocity magnitudes of the dark matter
particles in the velocity peaks on Earth.  The averages are over all
locations at the same distance (8.5 kpc) from the galactic center as we
are. 

In the course of this study, it was found that caustic rings
\cite{stw,sik} form in the coherent flows of dark matter particles 
associated with the velocity peaks.  A caustic is a location in physical 
space where the density is enhanced because the 3-dim. sheet on which the 
dark matter particles lie in 6-dim. phase space folds back there.  The
Zel'dovich ``pancake'' is an example of caustic.  It is a surface in 
physical space where the density diverges as ${1\over \sqrt{x}}$, $x$ being 
the distance to the surface.  The ring caustic is a closed line, or 
loop, where the density diverges as ${1\over x},~x$ being the distance 
to the line.  The line may however have internal structure.  This remains
to be investigated.  

A caustic ring is associated with each flow of dark matter particles in
and out of the central parts of the galaxy.  Thus, there is a caustic ring
due to particles falling through the galaxy for the first time, a caustic
ring of smaller size due to particles falling through for the second time,
a yet smaller ring due to particles falling through for the third time,
and so on.  For an arbitrary angular momentum distribution of the
infalling particles, the ring is a closed line of arbitrary shape. 
However, if the angular momentum distribution is dominated by a smooth
component which carries \underbar {net} angular momentum, the ring
resembles a circle.  If there is no angular momentum at all, the ring
reduces to a point at the galactic center.  The caustic rings get smeared
by the velocity dispersion of the infalling dark matter.  However, for a
galaxy like our own, that velocity dispersion has to reach a few 10 km/s
for the caustic rings to be completely washed out.  

The self-similar infall with angular momentum model of galactic halo
formation predicts the radii of the caustic rings in terms of the rotation
velocity $v_{rot}$ of the galaxy and the model parameters $\epsilon,
~j_{max}$ and $h$.  In theories of large scale structure formation with
cold dark matter and a Harrison-Zel'dovich spectrum of primordial density
perturbations, $\epsilon$ is in the range 0.2 to 0.3.  $j_{max}$ is the
maximum value of the dimensionless angular momentum distribution of the
infalling particles, as defined in ref. \cite{stw,sik}.  $h$ is the Hubble
constant in units of 100 km/s.Mpc.  It turns out that, in the range
$\epsilon \sim 0.2$ to 0.3, the ratios $a_n/a_{n-1}$ of successive ring
radii are nearly independent of $v_{rot},~\epsilon,~h$ and $j_{max}$. 
They have the values \beqn \{a_n/a_{n-1} : n=2,3,4,5...\} =
(0.49,~0.67,~0.76,~0.81,...)  \eeqn which are therefore predictions of the
self-similar infall with angular momentum model.  It is natural to expect
the caustic rings of a spiral galaxy to lie in its galactic plane.  In
that case, the caustic rings cause bumps in the galactic rotation curve
which are large enough to be observed.  It turns out that the rotation
curve of NGC3198, one of the best measured and often cited as providing
compelling evidence for halos of dark matter, has three bumps at radii
whose ratios match the first two entries on the RHS of Eq. (11).

\end{document}